\documentclass[]{spie}  

\usepackage{amsmath,amsfonts,amssymb}
\usepackage{graphicx}
\usepackage[colorlinks=true, allcolors=blue]{hyperref}

\title{Exoplanet Search and Characterization with the Proposed POET Canadian Space Mission}

\author[a]{*Stanimir Metchev}
\author[b]{Jason Rowe}
\author[c]{Paulo Miles-P\'aez}
\author[b]{Kelsey Hoffman}
\author[a]{Samantha Lambier}
\author[d]{Ryan Cloutier}
\author[a]{Hiroyuki Tako Ishikawa}
\author[e]{JJ Kavelaars}
\author[f]{Michelle Kunimoto}
\author[g]{David Lafreni\`ere}
\author[h]{Catherine Lovekin}
\author[a]{Eric Pilles}
\author[b]{John Ruan}
\author[a]{Jayshri Sabarinathan}
\author[i]{Gregg Wade}
\author[a]{Paul Wiegert}
\author[j]{Fr\'ederic Grandmont}
\author[j]{Anne-Sophie Poulin-Girard}
\author[k]{Simon Grocott}
\author[k]{Robert Zee}
\author[l]{Jean Dupuis}
\author[l]{Pierre Langlois}
\author[l]{Joel Roediger}

\affil[a]{The University of Western Ontario, Institute of Earth and Space Exploration, London, ON, Canada}
\affil[b]{Bishop's University, Sherbrooke, QC, Canada}
\affil[c]{Centro de Astrobiolog\'ia, CSIC-INTA, Madrid, Spain}
\affil[d]{McMaster University, Hamilton, ON, Canada}
\affil[e]{National Research Council -- Herzberg Astronomy and Astrophysics, Victoria, BC, Canada}
\affil[f]{The University of British Columbia, Vancouver, BC, Canada}
\affil[g]{Univertit\'e de Montr\'eal, Montr\'eal, QC, Canada}
\affil[h]{Mount Allison University, Sackville, NB, Canada}
\affil[i]{Royal Military College of Canada, Kingston, ON, Canada}
\affil[j]{ABB Measurement and Analytics, Qu\'ebec, QC, Canada}
\affil[k]{SFL Missions Inc., Toronto, ON, Canada}
\affil[l]{Canadian Space Agency, St-Hubert, QC, Canada}

\authorinfo{Further author information: (Send correspondence to S.M.)\\S.M.: E-mail: *smetchev@uwo.ca, Telephone: +1 519 661 2111, ext.\ 88438}

\begin{document} 
\maketitle

\begin{abstract}
The Photometric Observations of Exoplanet Transits (POET) is a proposed micro-satellite mission dedicated to the characterization and discovery of transiting exoplanets. POET has been identified as a top priority small-sat space mission in the Canadian Astronomy Long Range Plan 2020--2030. POET is being proposed as Canada's next astronomy space mission, with launch possible in late 2029.

POET is an iteration on the designs of the Canadian MOST and NEOSSat space missions, which had 15 cm-sized telescopes and observed only in the visible band pass. POET will have a larger 20 cm telescope aperture and three band passes: near-ultraviolet (nUV; 300--400 nm), visible near-infrared (VNIR; 400--900 nm), and short-wavelength infrared (SWIR; 900--1700 nm). All mission components either already have significant space heritage or are seeing rapid adoption in commercial space missions.

POET's simultaneous tri-band 300--1700 nm photometric monitoring will allow it to separate the impact of star spots on the transmission spectrum of extended atmospheres on super-Earth or larger exoplanets. 
POET's SWIR band is optimally sensitive to the emission peak of ultracool dwarf stars and would enable a systematic search for Earth-sized planets around them. POET aims to discover some of the nearest potentially habitable Earth-sized exoplanets that could be scrutinized for biosignatures with JWST or future telescopes. Herein we present the assembly of the POET Input Catalog of Ultracool Dwarfs and simulations of the expected yield of rocky planets with POET.
\end{abstract}


\section{INTRODUCTION}
\label{sec:intro} 

The Photometric Observations of Exoplanet Transits (POET)\cite{rowe_etal22} is a proposed micro-satellite mission dedicated to the characterization and discovery of transiting exoplanets, with launch possible in late 2029. POET has been identified as a top priority small-sat space mission in the Canadian Astronomy Long Range Plan 2020--2030\cite{barmby_etal21}. POET is being proposed as Canada's next astronomy space mission, through a funding model combining the Canadian Foundation for Innovation, the Canadian Space Agency, and the Canadian Nataional Research Council.

POET will have a 20 cm telescope aperture and three band passes: near-ultraviolet (nUV; 300--400 nm), visible near-infrared (VNIR; 400--900 nm), and short-wavelength infrared (SWIR; 900--1700 nm). The optomechanical assembly of POET has already been described in an earlier publication\cite{poulin-girard_etal24}, while its characteization in space-like conditions is the subject of a companion paper in these proceedings.\cite{poulin-girard_etal25}

POET has two principal science goals to be complete over a two-year main mission: 
\begin{enumerate}
    \item Exoplanet Characterization: the atmospheric characterization of known super-Earth or larger transiting planets. POET's simultaneous tri-band 300--1700 nm photometric monitoring will allow it to separate the impact of star spots on the transmission spectrum of extended atmospheres on super-Earth or larger exoplanets.\cite{rackham_etal18}
    \item Exoplanet Detection: the discovery of transiting potentially habitable rocky planets around the nearest ultracool dwarf-type stars. POET's SWIR band is optimally sensitive to the emission peak of ultracool dwarf stars and would enable a systematic search for Earth-sized planets around them. 
\end{enumerate}    

Additional science applications that take advantage of POET's simultaneous 300--1700 nm photometry and a polar orbit that allows a high ($>$80\%) duty cycle include solar system astronomy, asteroseismology, and transient follow-up.

This paper focuses on the second of the above two principal science goals: the discovery of rocky planets around ultracool dwarf stars. Ultracool dwarfs have effective temperatures ($T_{\rm eff}$) less than 2700 K and spectral types of M7 or later\cite{kirkpatrick_etal97}, including both very low-mass stars ($<$0.10 solar masses [$M_\odot$]) and substellar ($\lesssim$0.072$M_\odot$)\cite{chabrier_baraffe97} objects: brown dwarfs of spectral types L, T, or Y. These faintest of low-mass stars and the brown dwarfs remain beyond the photometric grasp of current all-sky transit exoplanet missions like the Transit Exoplanet Survey Satellite (TESS)\cite{ricker_etal15}. Any habitable-zone planets around these are expected to be in $<$1 week-long orbits around stars that are intrinsically among the dimmest. Because of the favorable planet-to-star flux ratios, any planets orbiting around ultracool dwarfs could offer some of the best opportunities for atmospheric characterization and biosignature detection with the James Webb Space Telescope (JWST) or with ground-based extremely large telescopes.


\section{TARGET SELECTION FOR ROCKY EXOPLANET DETECTION}
\label{sec:targets}

To discover Earth-sized planets via transits we aim to select a list of approximately 100--300 nearby ultracool dwarfs that maximize the probability for the detection of transiting rocky planets. Ultracool dwarfs have the smallest possible radii for main sequence stars: $\approx$0.1 solar radius or $\approx$10 Earth radii. This creates an $\approx$1\% flux deficit when an Earth-sized planet transits in front of it, which is generally trivial to detect. The ultracool dwarfs need only be bright enough to allow sufficient photometric precision to detect a 1\% transit with POET.

\subsection{Optimizing for Equator-on Viewing Geometry}

Rather than targeting a random sample of ultracool dwarfs, which would have only $\lesssim$3\% probability of transiting orbital geometry, we select ultracool dwarfs seen at their equator. As in our solar system, the orbits of most transiting exoplanets are aligned with the spins of their host stars\cite{winn_fabrycky15}. While spin-orbit {\sl mis-}alignment is not uncommon in systems with individual ice- or gas-giant planets\cite{ballard_johnson16}, nearly all systems with sub-Neptune planets, and multi-planet systems in particular, have small ($\le$20$^\circ$) obliquities\cite{winn_etal17}. Therefore, ultracool dwarfs viewed equator-on are among the best candidates to search for Earth-sized transiting planets. We hence seek ultracool dwarfs with spin axis inclinations of $i\gtrsim70^\circ$. 

While the spin axis inclination is not a direct observable, it can be inferred if the stellar radius $R$, rotation period $P$, and projected rotational velocity $v \sin(i)$ are known through:
\begin{equation}
\label{eq:inclination}
    \sin(i) = \frac{v\sin i}{P/(2\pi R)}.
\end{equation}
The period and projected rotational velocity can be determined observationally: through photometric monitoring and high-dispersion spectroscopy. If the distance to teh star is known, its radius can be obtained by comparing the total luminosity integrated over all wavelengths to a model of the photosphere that simultaneously fits for the effective temperature and the radius (see Equations \ref{eq:stellar_params} and \ref{eq:free_parameter}).

\subsection{Selection of the POET Input Catalog of Ultracool Dwarfs}

Our approach in building the input POET sample of ultracool dwarfs is as follows:
\begin{enumerate}
    \item Select a sample of candidate nearby ultracool dwarfs with known distances down to a brightness limit matched to the anticipated POET sensitivity. 
    \item Eliminate likely unresolved binary stars or stars with large radii (i.e., not dwarfs) by assessing their photometry. We refer to the list of candidate single stars as the POET Input Catalog of Ultracool Dwarfs.
    \item Obtain multi-wavelength information for all targets in the POET Input Catalog to build multi-wavelength spectral energy distributions.
    \item Determine the radii and effective temperatures of the ultracool dwarfs by fitting photospheric models to the spectra energy distributions, and using the known distances.
    \item Obtain the rotational periods of the ultracool dwarfs via photometric monitoring: either from existing all-sky synoptic surveys or from additional observations.
    \item Determine the projected rotational velocities of the ultracool dwarfs from archival or new high-dispersion spectroscopic observations.
\end{enumerate}


We detail our approach to each of the above steps in the following. Steps 1, 2, and 5 are similar to those described in our recent study of ultracool dwarf variables in TESS\cite{lambier_etal25}.

\begin{itemize}
    \item[1.] We have used Data Release 3 (DR3) of the Gaia mission\cite{gaia21} to select a flux-limited sample of ultracool dwarfs with known heliocentric distances. We use the 2MASS cross-identifications in DR3 to also extract 2MASS $J$-band magnitudes. The specific selection criteria are as follows:
    \begin{itemize}
        \item[a.] $J\leq14$ mag: anticipated sensitivity limit of POET in the SWIR\cite{rowe_etal22};
        \item[b.] Absolute $G_{RP}\geq10.2$ mag (and when no $G_{RP}$ mag in DR3, absolute $G\geq14$ mag): selecting intrinsically faint stars, such as ultracool dwarfs;
        \item[c.] $G-G_{RP}\geq 1.4$ mag: 	selecting candidate ultracool ($\geq$M6) dwarfs by their red optical colors; 
        \item[d.] $G-J\geq 0.85 (G-G_{RP})^2 - 0.95(G-G_{RP}) + 2.9$ mag: selecting candidate ultracool dwarfs by their red optical minus SWIR colors;
        \item[e.] SNR($G$) $\geq$ 20: selecting high-quality Gaia photometry;
        \item[f.] SNR($G_{RP}$) $\geq$ 10: selecting high-quality Gaia photometry; 
        \item[g.] corrected $G_{BP}/G_{RP}$ excess factor $\leq 1.55 + 0.07(G_{BP} - G_{RP})^2$: selecting high-quality Gaia photometry; 
        \item[h.] relative parallax error $\leq$ 0.2: selecting high-quality Gaia parallaxes.
    \end{itemize}
\end{itemize}

\begin{figure}
    \centering
    \includegraphics{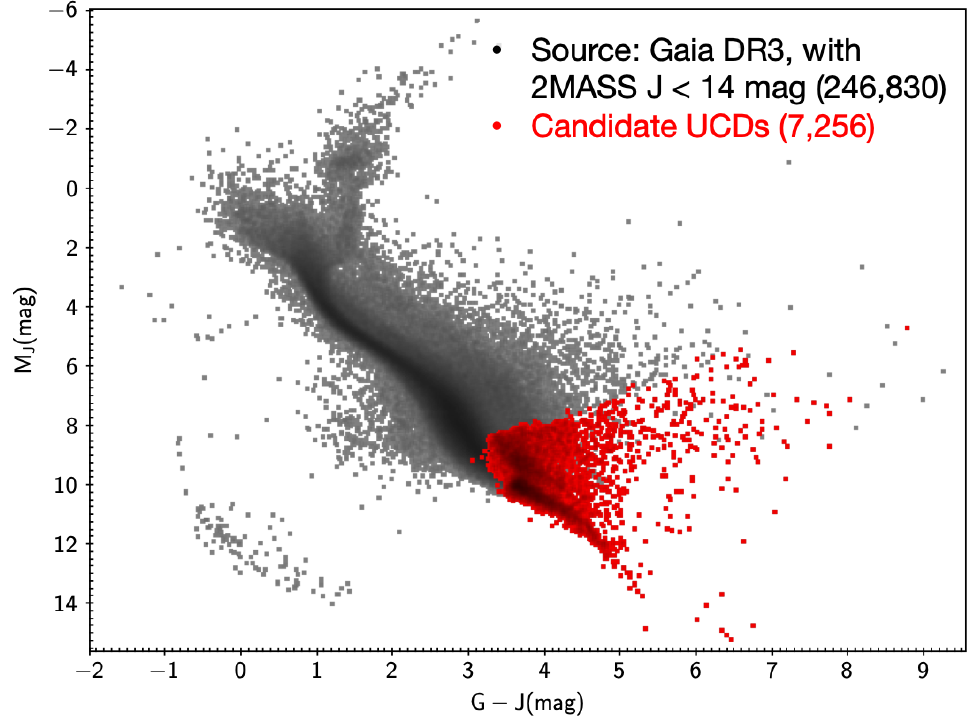}
    \caption[Selection of the ultracool dwarf targets for POET]{Selection of the POET Input Catalog of Ultracool Dwarfs via Gaia DR3 and 2MASS $J$-band photometry and parallax.}
    \label{fig:ucd_catalog}
\end{figure}

The above criteria produced a list of 7,256 candidate ultracool dwarfs for the POET Input Catalog. These are shown with red dots on a color-absolute magnitude in Figure \ref{fig:ucd_catalog}.

\begin{figure}
    \centering
    \includegraphics[width=0.8\textwidth]{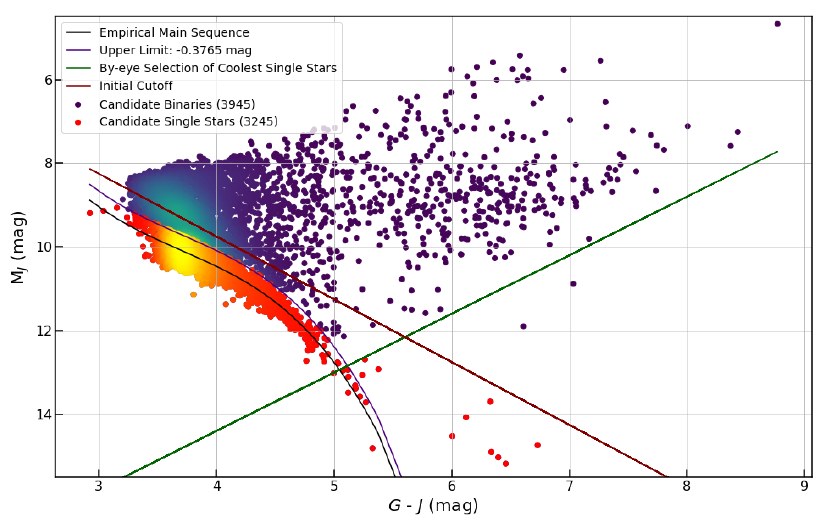}
    \caption[Ultracool dwarf targets, excluding candidate binary stars]{Elimination of candidate binary stars from the initial sample by removing stars 0.38 mag above an empirical fit to the main sequence. The resulting sample of 3245 candidate single stars (red) is the POET Input Catalog of Ultracool Dwarfs.}
    \label{fig:ucd_catalog_no_binaries}
\end{figure}

\begin{itemize}
    \item[2.] We exclude binary stars or over-luminous stars with large radii from our candidate sample, as their over-luminosity hinders the detection of Earth-sized planets. We implement two approaches to eliminate these. We first check whether a candidate ultracool dwarf is brighter by $\geq$0.38 mag in absolute $J$ magnitude from its expected main-sequence location.  This could indicate either unresolved binarity, or a large radius: e.g., because the star is young, and still contracting, or because it is a sub-giant, rather than a dwarf. Our final list of candidate single ultracool dwarf stars contains 3245 objects, shown in red in Figure \ref{fig:ucd_catalog_no_binaries}. These objects constitute the POET Input Catalog of Ultracool Dwarfs.
\end{itemize}

\begin{figure}
    \centering
    \includegraphics[width=0.7\textwidth]{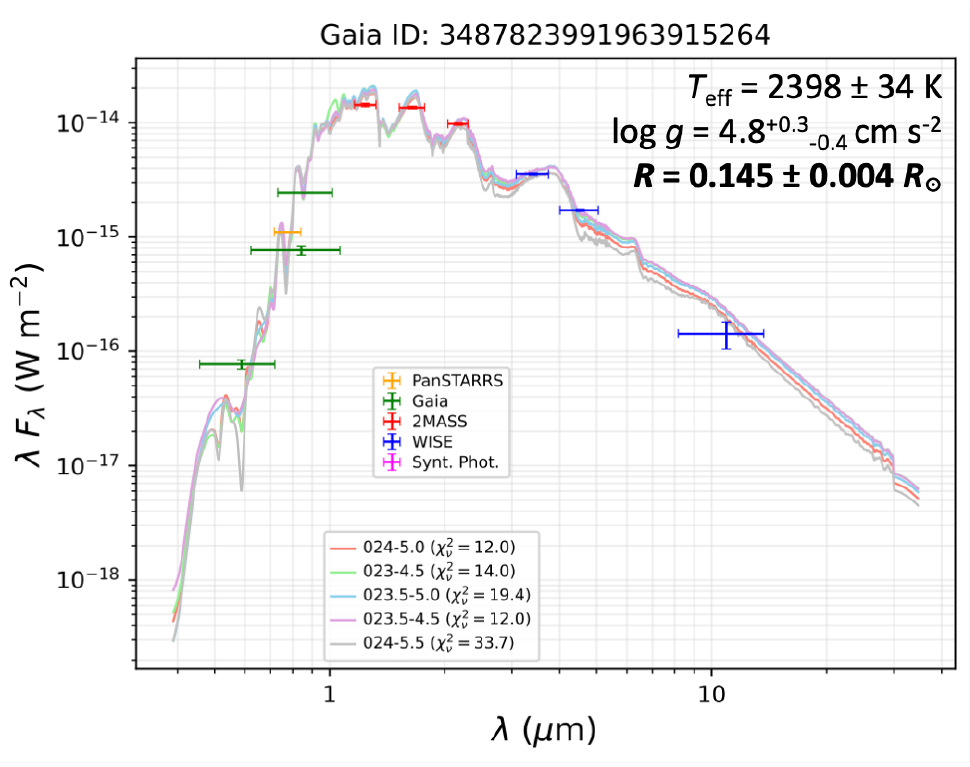}
    \caption[Determination of best-fit stellar parameters]{Determination of best-fit stellar parameters (effective temperature $T_{\rm eff}$, surface gravity $g$, stellar radius $R$) from a photospheric model fit to multi-wavelength photometry.\cite{suarez_etal25}}
    \label{fig:ucd_stellar_params}
\end{figure}

\begin{itemize}
    \item[3.] We cross-matched our Gaia DR3 candidate ultracool dwarfs with other all-sky or large area catalogs, including 2MASS\cite{skrutskie_etal06}, AllWISE\cite{cutri_etal13}, Pan-STARRS\cite{kaiser_etal02} and SDSS\cite{york_etal00} to obtain photometric coverage from 400 nm to 10 $\mu$m or 20 $\mu$m. In up to 15\% of cases, especially in AllWISE, the photometry was contaminated by nearby sources. In some cases we were able to extract photometry from a higher-resolution survey (e.g., unWISE\cite{schlafly_etal19} vs.\ AllWISE or UKIDSS\cite{lawrence_etal07} vs.\ 2MASS), and added that to the spectral energy distribution. In cases where such is not available, we ignore the contaminated photometry in our subsequent analysis.
    \item[4.] We ingest the spectral energy distribution photometry and the astrometric distance to each star, to compare against photospheric models, and to determine best-fit stellar radius and effective temperature\cite{suarez_etal25}. We use a chi-squared metric in determining the best-fit parameters:
    \begin{equation}
    \label{eq:stellar_params}
        \chi^2 = \frac{1}{N-2}\sum_{i=1}^{N}\left[\frac{(O_i - MY_i)^2}{\sigma^2_i + \sigma_M^2}\right],
    \end{equation}
    where $N$ is the number of photometric measurements, $O_i$ are the observed fluxes, $Y_i$ are the fluxes predicted by the model, $\sigma_i$ and $\sigma_M$ are random and systematic uncertainties, and $M$ is a free scaling parameter. In practice, 
    \begin{equation}
    \label{eq:free_parameter}
        M = \left(\frac{R}{D}\right)^2,
    \end{equation}
    where $D$ is the known (from Gaia DR3) heliocentric distance. Hence, the chi-squared minimization yields the radius $R$, whereas the best-fit model gives the effective temperature $T_{\rm eff}$. Example output is shown in Figure \ref{fig:ucd_stellar_params}.
\end{itemize}

\begin{figure}
    \centering
    \includegraphics{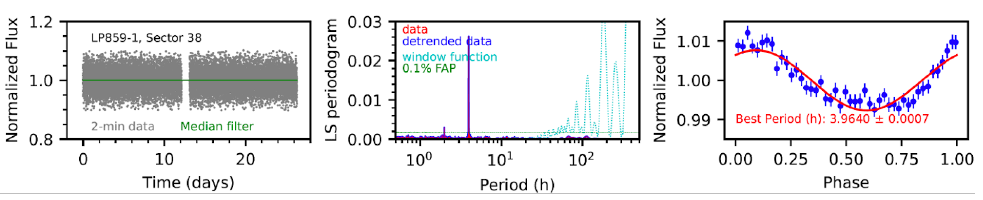}
    \caption[Discovery of a 3.96-day period in low-SNR data from TESS]{Discovery of a 3.96-day period in low-SNR data from TESS of the $T = 14.4$ mag M7 dwarf LP 869-1\cite{miles-paez_etal23}.}
    \label{fig:ucd_tess}
\end{figure}

\begin{itemize}
    \item[5.] About 85\% of the POET Input Catalog Ultracool Dwarfs lie within the footprint of the NASA TESS mission. Even if the majority of our Input Catalog is fainter than the $T\sim 14$ mag limit considered viable for transit exoplanet detection with TESS, the TESS data are still useful to detect highly repeating photometric periodicities caused by stars spots and stellar rotation. We have developed a custom code to detect photometric periodicities in such low-SNR data\cite{miles-paez_etal23, lambier_etal25}. An example of the success of our approach is shown in Figure \ref{fig:ucd_tess}, which reveals a 3.96 day period in a $T = 14.4$ mag M7-type ultracool dwarf. Preliminary findings from TESS and ground-based photometric observations have already been published\cite{miles-paez_etal23, lambier_etal25}.
    \item[] For the remaining $\approx$15\% fraction of POET Input Catalog Ultracool Dwarfs, for which no TESS (or Kepler\cite{boricki_etal10}, or K2) monitoring exists, we are undertaking an observing campaign with Canada's Near-Earth Object Survey Satellite (NEOSSat\cite{wallace_etal04}) or with ground-based telescopes. The campaign is expected to run until 2028.
    \item[] We anticipate to be able to obtain photometric periods for $\sim$30\% of our POET Input Catalog ultracool dwarfs. The main culprit for the low success rate would be the viewing geometry: stars seen closer to pole-on will not present rotationally modulated flux variations, even if they do have surface spots. At the same time, the $\sim$70\% of the Input Catalog that does not yield photometric periods will not be of interest for transit planet detections with POET precisely because of the unfavorable viewing geometry.
\end{itemize}

\begin{itemize}
    \item[6.] We are in parallel conducting a large campaign of high-dispersion ($R\equiv\lambda/\Delta\lambda >20,000$) spectroscopic observations, where such data are available from the literature, of POET Input Catalog targets using 3--8 meter-class ground-based telescopes. The high-dispersion spectroscopy allows us to measure the projected rotational velocity $v \sin(i)$. The spectroscopic data gathering is also anticipated to extend until 2028. 
\end{itemize}

Once all photometric rotation period and spectroscopic rotational velocity data are compiled, we will prioritize a sample of 100--300 ultracool dwarfs viewed nearly at the equator for the POET rocky planet transit search (Section~\ref{sec:yield}).

\section{SIMULATIONS OF ROCKY PLANET DISCOVERIES WITH POET}

We are developing an end-to-end simulation of the rocky exoplanet discovery component of POET operations. This  will serve to validate and optimise the observing sample and approach, and to allow flexibility in modifying the observing plan depending on changes in mission capabilities. The science validation testing entails simulating the potential outcomes of the exoplanet detection survey based on varying mission parameters: use of the VNIR vs.\ the SWIR detector on POET, duty cycle, observing duration, minimum observing block. Our simulation steps include:
\begin{enumerate}
    \item Assessing the relative sensitivity to ultracool dwarfs and transiting planet radii with the VNIR and the SWIR detectors.
    \item Projecting the success rate in detecting Earth-sized exoplanets with various combinations of target sample (size, brightness distribution) and POET mission parameters.
    \item Projecting and optimising the POET mission exoplanet discovery yield for various POET realisations.
\end{enumerate}


These tasks are described in Sections~\ref{sec:sensitivity}--\ref{sec:yield}.

\subsection{Transit Detection Sensitivity Projections}
\label{sec:sensitivity}

Initial simulations of the sensitivity to transiting planets around ultracool dwarfs revealed that POET would be sensitive to one Earth-radius transiting planets around ultracool dwarfs brighter than $I = 13$ mag or $J = 13$ mag\cite{rowe_etal22}. 
Since ultracool dwarf colors are redder than $I-J = 2.6$ mag, this strongly favors $J$-band (SWIR) over $I$-band (VNIR) observations. There would be over twice as many accessible ultracool dwarfs to POET in the SWIR. 


However, the intrinsic noise of SWIR detectors is higher than for VNIR detectors. We have therefore pursued an assessment of POET's SWIR and VNIR performance with realistic sensitivity estimates. We have adopted empirical estimates for the various noise parameters, informed from detectors used in prior space missions (CCD 47-20 sensor\cite{walker_etal03} for VNIR) or ground-based telescopes (a FLIR InGaAs camera\cite{simcoe_etal19} for SWIR), as follows:
    \begin{itemize}
        \item VNIR: read noise = 8 e$^-$, dark current = 0.4 e$^-$/sec/pix;
        \item SWIR: read noise = 43 e$^-$, dark current = 113 e$^-$/sec/pix.
    \end{itemize}

We further assume 20\% overall throughput between 400--900 nm (VNIR) and 1100--1700 nm (SWIR). We use BT-Settl photospheric models\cite{allard_etal12} to estimate the S/N of a 3 min-long long exposure for ultracool dwarfs in the $8<G<20$ mag range. At this stage we have used only a $T_{\rm eff} = 2500$ K and a $T_{\rm eff} = 1500$ K model, representative of M8 and L6 dwarfs, and have adopted the closest one in temperature for ultracool dwarfs of other spectral types.

\begin{figure}
    \centering
    \includegraphics[width=0.7\textwidth]{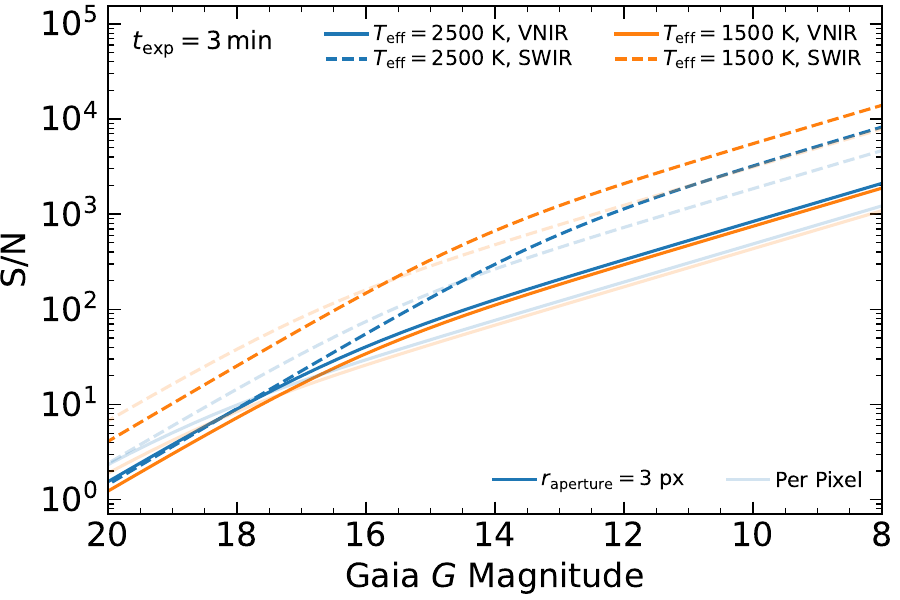}
    \caption[Comparative performance in S/N as a function of Gaia G magnitude]{Comparative performance in the attained S/N on $T_{\rm eff}$ = 2500 K (blue; spectral type M8) and $T_{\rm eff}$ = 1500 K (orange; spectral type L6) dwarfs for SWIR (dashed lines) and VNIR (solid lines) as a function of Gaia $G$ magnitude. The SWIR is advantageous for brighter ($G<16$ mag) objects by up to a factor of 8 for the colder (1500 K) ultracool dwarfs, but does not offer advantages on the warmer (2500 K) ultracool dwarfs at faint magnitudes.
    }
    \label{fig:ucd_detector_performance}
\end{figure}

The sensitivity comparison is shown Figure \ref{fig:ucd_detector_performance}. It demonstrates factors of 3 to 8 higher S/N in the SWIR (dashed orange line) compared to the VNIR (solid orange line) for the colder (1500 K) ultracool dwarfs. For the warmer (2500 K) stars there is an S/N advantage in the SWIR (dashed blue line) only for stars brighter than $G\approx 16$ mag. At fainter $G>16$ magnitudes the SWIR and the VNIR (solid blue line) are equally dominated by photon noise. 

\subsection{Detectability of Transiting Rocky Planets}

We use the sensitivity projections from Figure \ref{fig:ucd_detector_performance} and the magnitude distribution of ultracool dwarfs in the POET Input Catalog to perform an initial assessment of the relative detectability of Earth-sized rocky planets with POET. For the purpose, we initially assume an uninterrupted 3-day stare on each star, and we simulate each observation 1000 times. We assume that each star has intrinsic and periodic stellar variability with amplitude between 0.001 and 0.05 mag and period between 0.5 and 40 h, as currently known of ultracool dwarfs\cite{miles-paez_etal23}. We injected a single transit in a random location of each simulated light curve. (That is, at this stage we assume that each star has a planet, and that it happened to transit during the observation!) 
We used planet radii between 0.6 and 6.6 Earth's radius following the radius distribution of planets around dwarfs cooler than 4000 K\cite{dressing_charbonneau15}. We then ran a Box Least Squares algorithm\cite{kovacs_etal02} that searches for the transit, and consider the transit detected if the retrieved transit depth, duration, and central time are within 5\% of the injected values.

\begin{figure}
    \centering
    \includegraphics[height=6.5in]{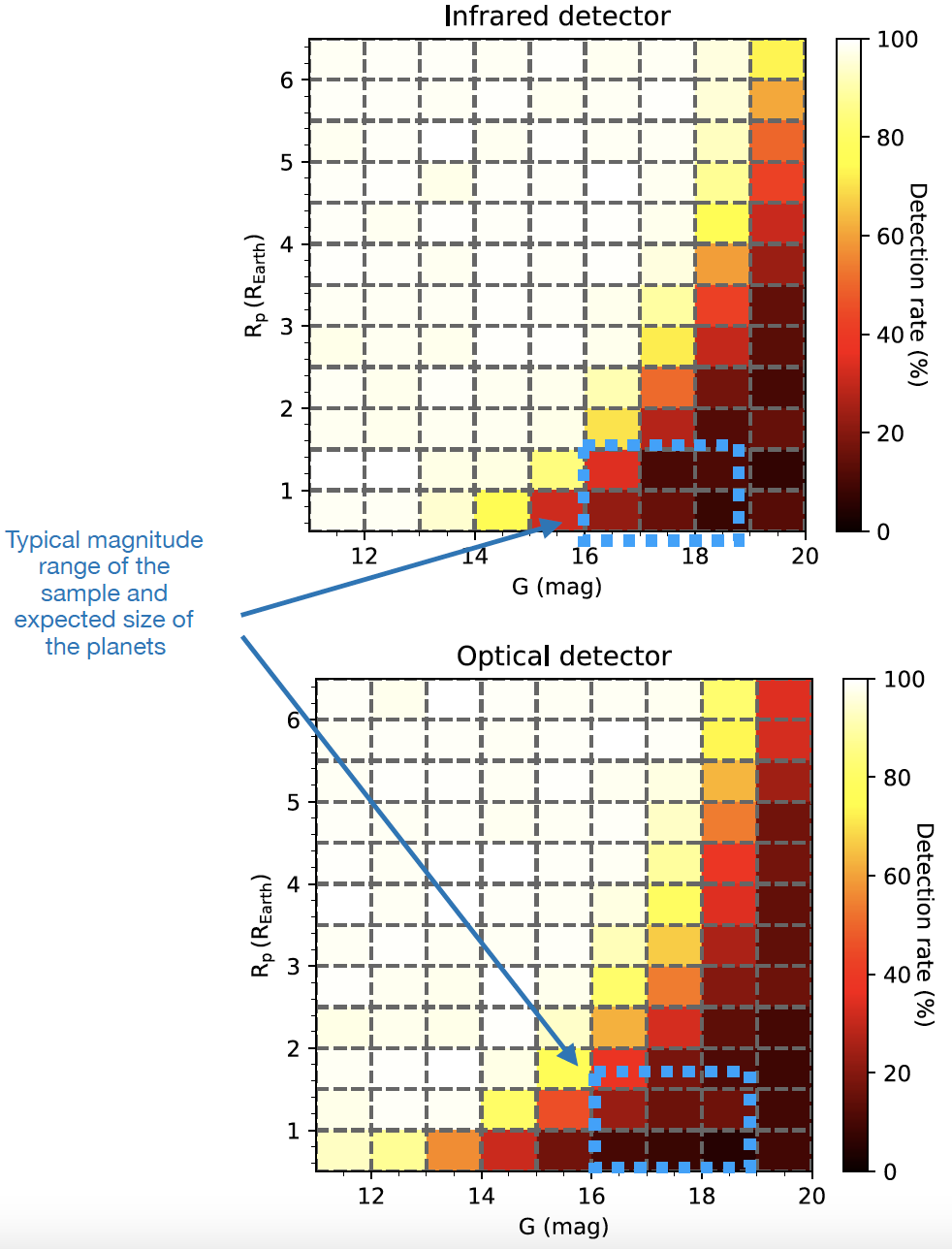} 
    \caption[Simulated detection rates of rocky planets with an SWIR) and a VNIR detector]{Simulated detection rates of rocky planets with an infrared (SWIR; upper panel) and an optical (VNIR; lower panel) detector on POET. 
    The typical magnitude range ($16<G<19$ mag) of the target sample (Figure~\ref{fig:UCD_hists}) the expected size range (0.5--1.5 Earth radii\cite{ment_charbonneau23}) of rocky planets around ultracool dwarfs are enclosed in the dashed rectangle on each panel.
    An SWIR detector has an overall factor of $\approx$2 advantage in detection rates at these small planet sizes and faint optical magnitudes.}
    \label{fig:ucd_sim_detection_rates}
\end{figure}

\begin{figure}[ht]
\centering
\begin{tabular}{c}
\includegraphics[height=0.4\textwidth]{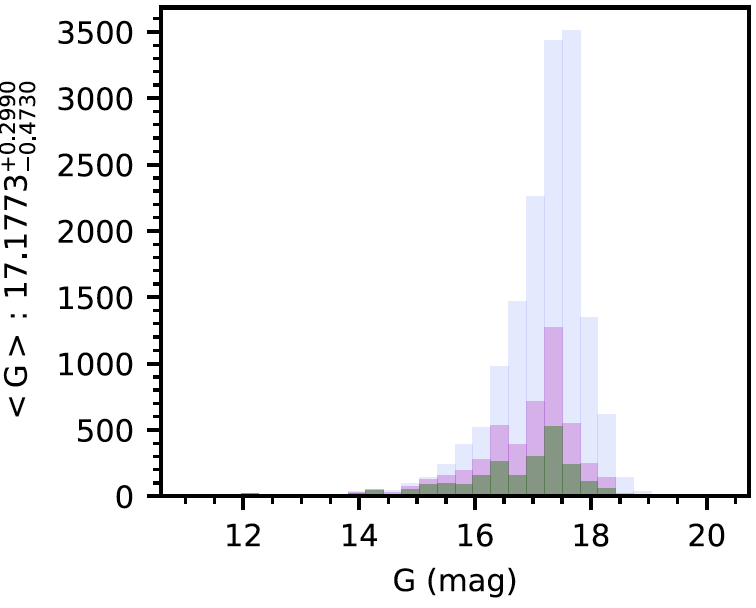}
\includegraphics[height=0.4\textwidth]{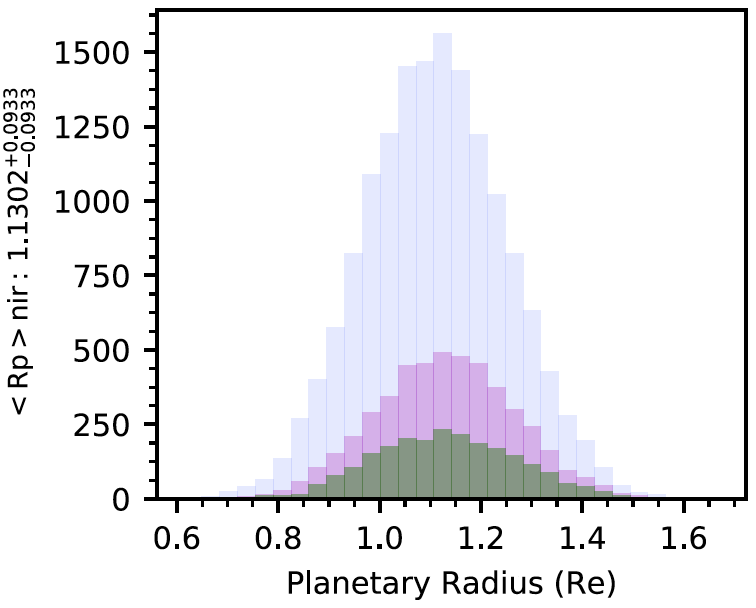}
\end{tabular}
\caption[Distributions of Host Star Magnitudes and Exoplanet Radii]{Distribution of host-star magnitudes (\textit{left}) and exoplanet radii (\textit{right}, in units of Earth's radius) in the survey simulations for rocky planet detection around ultracool dwarfs. Light blue represents all simulated transiting planets that fall in the observing window of their stars. Magenta and green represent the subsets of planets for which the transits would be successfully recovered given POET sensitivity considerations for the SWIR and the VNIR, respectively. The mean magnitude of the stellar host is $G = 17.17$ mag. The mean radius of the detected planet is 1.13 Earth radii. The detection rates in the SWIR are $\approx$2 times higher than in the VNIR.}
\label{fig:UCD_hists}
\end{figure}

Figure \ref{fig:ucd_sim_detection_rates} details these preliminary findings. It shows that for large ($>$2.5 Earth radii) planets both the SWIR and the VNIR have a very high success rate (70\%--100\%) of detection on bright ($G<17$ mag) stars. The advantage of SWIR observations is seen toward fainter stars and smaller planets, 
which are also the objective of the POET Exoplanet Detection survey. Figure~\ref{fig:UCD_hists} further shows the distribution of host star parameters and exoplanet radii for simulated and detected exoplanets. Both Figures \ref{fig:ucd_sim_detection_rates} and \ref{fig:UCD_hists} illustrate that a larger fraction of the population of the Earth-sized planets are detectable in the SWIR.

As a cautionary note, these estimates are still preliminary. In particular, the VNIR noise parameters have yet to be finalized through comparisons of similar CCD47-20 detectors on previous missions (MOST\cite{walker_etal03}, CHEOPS\cite{benz_etal21}), and current current manufacturer projections. The noise parameters of the SWIR detector from FLIR are also based solely on ground-based measurements\cite{simcoe_etal19}. A different SWIR camera, the Owl 1280 camera from Raptor Photonics, is under consideration for POET. The Owl 1280 SWIR camera already have space heritage in the private space sector. Its characterization for astronomical observations is currently on-going. Follow-up work on the simulated detection rates will include updated noise expectations for the VNIR and SWIR bands.

\subsection{Anticipated Yield}
\label{sec:yield}

The overall planet yield from the ultracool dwarf survey for rocky planets is strongly dependent on the underlying exoplanet population around very low-mass stars and brown dwarfs, the parameters of which are still largely unknown. Estimates for the occurrence rate of Earth-sized planets in short-period ($<$7--50 day) around low-mass stars are consistently high, and in the 50\%-60\% range\cite{dressing_charbonneau15, ment_charbonneau23}. More uncertain is the planetary size distribution. There is an increasing occurrence rate toward smaller rocky planets around early-M dwarfs\cite{dressing_charbonneau15}, as might be expected from the broader planet size-frequency distribution\cite{howard13}. However, toward mid- to late-M dwarfs, the size distribution for rocky planets may be narrower, centred on 1.14 Earth radii with a standard deviation of 0.11 Earth radii\cite{ment_charbonneau23}. In principle, the latter\cite{ment_charbonneau23} result is more appropriate for the POET ultracool dwarf targets. However, it is also based on a much smaller sample and number of detections than the early-M dwarf analysis\cite{dressing_charbonneau15}. We conservatively adopt the more stringent parent sample statistics\cite{ment_charbonneau23} in our simulations, but anticipate that they may under-estimate the true yield of rocky planets around ultracool dwarfs by a factor of at least 3, since they severely under-represent the more easily detectable larger planets. Ultimately, POET will be able to determine both the occurrence rate and the size distribution of rocky planets around ultracool dwarfs with much better certainty.

The rocky planet yield from POET is also dependent on the total observing time available for the Exoplanet Detection campaign, the POET duty cycle, the choice of detector (and sensitivity; Section~\ref{sec:sensitivity}), the number of stars chosen in the survey sample, and the range of spin axis inclinations of the candidate host star sample. 
    
We assume that the total Exoplanet Detection campaign is fixed at 220 days, which accounts for a year-long survey with an 80\% duty cycle and additional 25\% of time set aside for General Observer programs. A 90\% duty cycle and no General Observer program would increase yields by 50\%, compared to the estimates below. Within the time constraints of the 220 days, we test observing strategies that range from shorter observations (0.25 day stares) of a very large number (880) of targets to longer observations (7 day stares) of a much smaller (31) sample. We further test three assumptions for lower limits on the inclination distribution of the parent sample: $i>60^\circ$, $i>70^\circ$, and $i>75^\circ$. For each of these survey parameter combinations we perform 5000 Monte Carlo simulations to assess the exoplanet yield.
    
\begin{figure}[ht]
\centering
\includegraphics[width=0.8\textwidth]{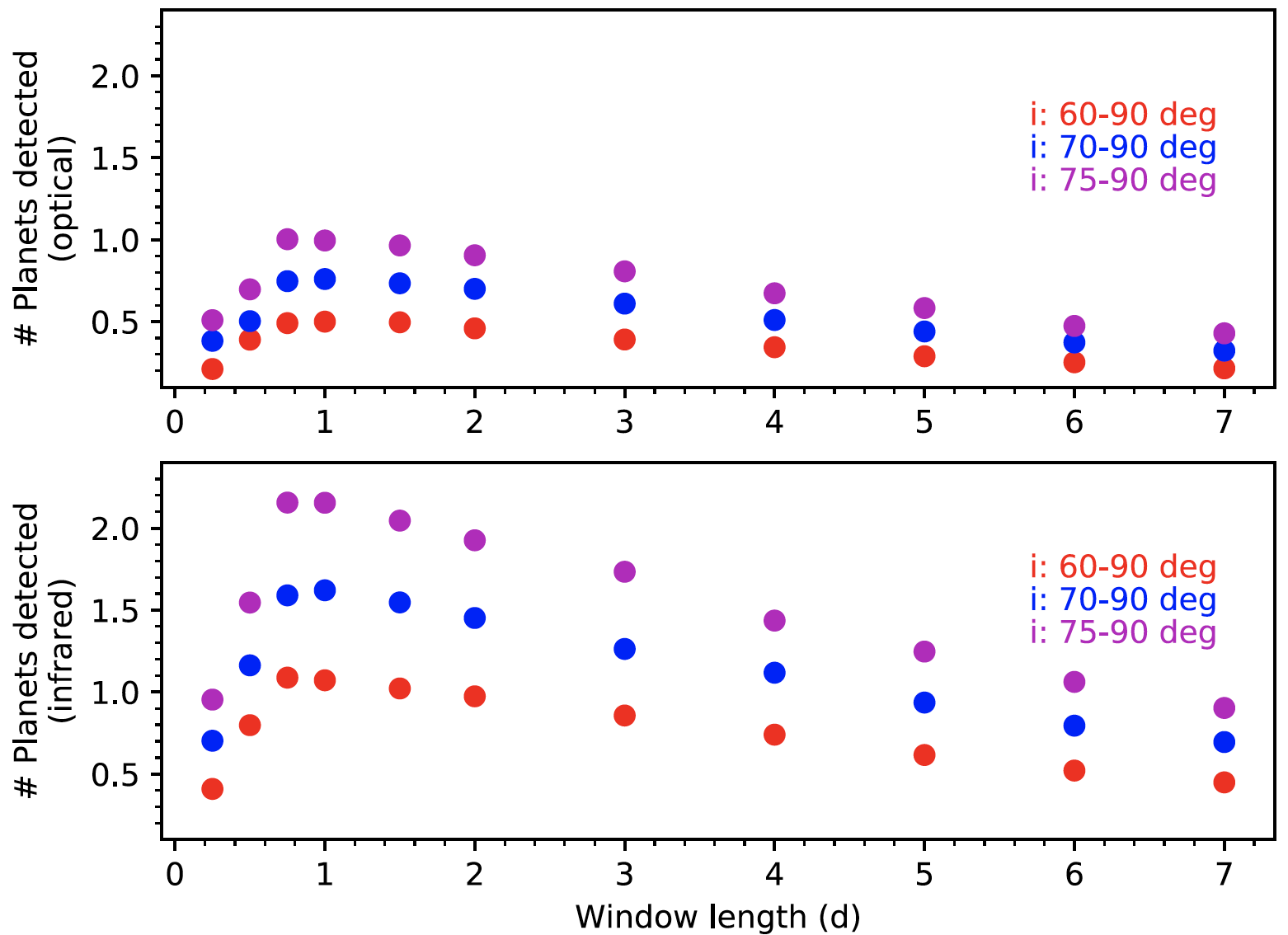}
\caption[Anticipated Rocky Planet Yield from Exoplanet Detection Survey of ultracool Dwarfs]{Anticipated yield of rocky planets with a VNIR (\textit{top panel}) or an SWIR (\textit{bottom panel}) detector. The yields are based on conservative assumptions for the size distribution of rocky planets (peaking at 1.1 Earth radii\cite{ment_charbonneau23}) and for the duration of the Exoplanet Detection survey (220 days, equivalent to a year at 80\% duty cycle and 25\% of time set aside for general observer programs). The detection rates in the SWIR are $\approx$2 times higher than in the VNIR. The selection of stars viewed closer to equator-on ($i>75^\circ$ inclinations) significantly improves the yield. More optimistic assumptions, based on size distributions that include larger rocky planets\cite{dressing_charbonneau15}, more efficient duty cycle (90\%), or less general observer time would raise the projected yields by a factor of at least 3.}
\label{fig:UCD_yields}
\end{figure}

Figure \ref{fig:UCD_yields} shows results under the most conservative of assumptions: the narrow planet size distribution centred on small (1.1 Earth) planetary radii\cite{ment_charbonneau23}, and a 220-day Exoplanet Discovery campaign (80\% duty cycle and additional 25\% of time set aside for General Observer programs). Figure \ref{fig:UCD_yields} demonstrates the importance of using an SWIR over a VNIR detector (factor of 2 increase in yield), and of careful sample selection (factor of 2 increase in yields for $i>75^\circ$ vs.\ $i>60^\circ$ sample). The Figure also shows that the yield is fairly constant for stares between 0.75--2 days in duration, which corresponds to sample sizes between 293 and 110 stars observed over one year. This is an important outcome, as it allows greater flexibility in potentially focusing on a smaller number of higher-inclination targets without sacrificing yield.

The upshot from the Exoplanet Detection survey simulations is that under conservative assumptions the yield from POET would be 2 rocky planets in short-period ($<$7-day) orbits around some of the nearest stars. These rock planets would be among the most amenable for atmospheric characterization with JWST. Relaxing the assumptions increases the anticipated yield by a factor of up to 3, suggesting that a year-long dedicated use of POET for a transiting rocky planet survey could yield up to 6 new ultracool dwarf hosts of Earth-sized planets. The number of new exoplanet discoveries could further at least double if the planets themselves are in multi-planet systems, as is common for low-mass stars. Indeed, the cumulative planet occurrence rate of rocky planets with radii between 1 and 4 times Earth's and periods shorter than 200 days is estimated at $2.5\pm0.2$ planets per early-M dwarf\cite{dressing_charbonneau15}. 
    
A broader range of estimates for the yield of the Exoplanet Detection campaign are planned for future studies, as the complexity and predictive power of the survey simulations are augmented.

\section{DISCUSSION and CONCLUSIONS}

We have described the assembly of the $>$3000-star POET Input Catalog of Ultracool Dwarfs, from which we will downselect between 100--300 top-priority targets for a year-long survey for transiting rocky planets with the POET space mission. We have performed initial sensitivity estimates and simulations (Figure \ref{fig:ucd_sim_detection_rates}--\ref{fig:UCD_yields}) of the detectability of rocky planets around ultracool dwarfs, and have estimated the yield of POET's year-long Exoplanet Detection campaign. The presented simulations are based on conservative assumptions for the size distribution of rocky planets and the time dedicated to a planet survey. Nonetheless, they point to an enhanced planet yield---by 100\%---in POET's SWIR channel over the VNIR channel.

At a minimum, we anticipate the discovery of two new rocky planets around the nearest stars with POET in the SWIR. Under potentially more realistic assumptions and a higher fraction of POET time dedicated to the Exoplanet Detection survey, the number of new rocky planet discoveries could be at least 6, and potentially well over 10 given the high degree of multiplicity in rocky planet systems.

More generally, the dedicated Exoplanet Detection survey with POET will for the first time establish the frequency, size, and orbital distribution of rocky planets around the dimmest and lowest-mass stellar objects. These are high-priority targets for exoplanet characterization, and yet have been largely unexplored because of their intrinsic faintness.

Newly-discovered Earth-sized planets around the nearest ultracool dwarfs would be excellent targets for atmospheric characterization. With orbital periods $<$7 days, these are likely to be in the habitable zones of their ultracool host stars. They would automatically become top-priority targets for biosignature gas searches with the Webb Space Telescope, or with the Habitable World Observatory further in the future. Hence, POET could deliver some of the most promising Earth analogues for the search for extrasolar life.



\acknowledgments 
 
This program is undertaken with the financial support of the Canadian Space Agency via a Science Maturation Study contract (No.\ 9F050-170207/003/MTB), a Space Technology Development Program contract (No.\ 9F063-190729/002/MTB), and three Flights for Advancement of Space Technology grants (Ref.\ No.\ 18FAWESC13, 19FAWESB40, 21FAUWOB12).

\bibliography{bibliography}
\bibliographystyle{spiebib} 

\end{document}